\documentclass{PoS}

\title{Static-light meson masses from twisted mass lattice QCD}

\ShortTitle{Static-light meson masses from twisted mass lattice QCD}

\author{Karl Jansen \\
        DESY, Platanenallee 6, D-15738 Zeuthen, Germany \\
        E-mail: \email{karl.jansen@desy.de}}

\author{Chris Michael, Andrea Shindler \\
        Theoretical Physics Division, Department of Mathematical Sciences, University of Liverpool, Liverpool L69 3BX, UK \\
        E-mail: \email{c.michael@liverpool.ac.uk} \\
        E-mail: \email{andrea.shindler@liverpool.ac.uk}}

\author{\speaker{Marc Wagner}\thanks{On behalf of the ETM Collaboration.} \\
        Humboldt-Universit\"at zu Berlin, Institut f\"ur Physik, Newtonstra{\ss}e 15, D-12489 Berlin, Germany \\
        E-mail: \email{mcwagner@physik.hu-berlin.de}}

\abstract{
We compute the static-light meson spectrum using two-flavor Wilson twisted mass lattice QCD. We have considered five different values for the light quark mass corresponding to $300 \, \textrm{MeV} \ltapprox m_\mathrm{PS} \ltapprox 600 \, \textrm{MeV}$. We have extrapolated our results, to make predictions regarding the spectrum of $B$ and $B_s$ mesons.
}

\FullConference{The XXVI International Symposium on Lattice Field Theory \\
                 July 14 - 19, 2008\\
                 Williamsburg, Virginia, USA}

\newcommand{\ltapprox}{\raisebox{-0.5ex}{$\,\stackrel{<}{\scriptstyle\sim}\,$}}

\begin{document}

% ********************
% ********************
% ********************
% ********************
% ********************

\section{Introduction}

A systematic way to study $B$ and $B_s$ mesons from first principles is lattice QCD. Since $a m_b > 1$ at currently available lattice spacings for large volume simulations, one way to proceed is to use for the $b$ quark a formalism such as Heavy Quark Effective Theory (HQET). In this paper we consider the leading order of HQET, which is just the static limit. In this limit a ``$B$ meson'' will be the ``hydrogen atom'' of QCD. States are either labeled by $J^\mathcal{P} = (L \pm 1/2)^\mathcal{P}$, where $L$ denotes orbital angular momentum, $\pm 1/2$ the coupling of the light quark spin and $\mathcal{P}$ parity, or by $S \equiv (1/2)^-$, \\ $P_- \equiv (1/2)^+$, $P_+ \equiv (3/2)^+$, $D_- \equiv (3/2)^-$, ... In the static limit these states will be doubly degenerate, since there is no interaction with the heavy quark spin, i.e.\ the total spin for given $J^\mathcal{P}$ is $F^\mathcal{P} = J^\mathcal{P} \pm 1/2$.

The spectrum of static-light mesons has been studied comprehensively by lattice methods in the quenched  approximation with a rather coarse lattice spacing \cite{Michael:1998sg}. More refined lattice studies with $N_f=2$ flavors of dynamical sea quarks have also explored this spectrum \cite{Green:2003zza,Burch:2006mb,Koponen:2007fe,Foley:2007ui,Burch:2007xy}. Here we also use $N_f=2$ and are able to reach lighter sea-quark masses, which are closer to the physical $u/d$ quark mass allowing more reliable extrapolations.

% ********************
% ********************
% ********************
% ********************
% ********************
% ********************

\section{Static-light meson creation operators}

In the continuum an operator creating a static-light meson with well defined quantum numbers $J^\mathcal{P}$ is given by
\begin{eqnarray}
\label{EQN001} \mathcal{O}^{(J^\mathcal{P})}(\mathbf{x}) \ \ = \ \ \bar{Q}(\mathbf{x}) \int d\hat{\mathbf{n}} \, \Gamma^{(J^\mathcal{P})}(\hat{\mathbf{n}}) U(\mathbf{x};\mathbf{x}+r \hat{\mathbf{n}}) \psi(\mathbf{x}+r \hat{\mathbf{n}}) .
\end{eqnarray}
$\bar{Q}$ creates an infinitely heavy antiquark, $\int d\hat{\mathbf{n}}$ denotes an integration over the unit sphere, $U$ is a spatial parallel transporter, and $\psi$ creates a light quark separated by a distance $r$ from the antiquark. $\Gamma^{(J^\mathcal{P})}$ is an appropriate combination of spherical harmonics and $\gamma$-matrices yielding a well defined total angular momentum $J$ and parity $\mathcal{P}$. The meson creation operators we use in this paper are listed in Table~\ref{TAB001}.

\begin{table}[h!]
\begin{center}

\begin{tabular}{|lc|c|c|c|c|}
\hline
 & & & & & \vspace{-0.40cm} \\
\multicolumn{2}{|c|}{$J^\mathcal{P}$} & $F^\mathcal{P}$ & $\Gamma^{(J^\mathcal{P})}(\mathbf{x})$ & $\mathrm{O}_\mathrm{h}$ & lattice $J^\mathcal{P}$ \\
 & & & & & \vspace{-0.40cm} \\
\hline
 & & & & & \vspace{-0.40cm} \\
$(1/2)^-$ & $[S]$ & $0^-,1^-$ & $\gamma_5 \ \ , \ \ \gamma_5 \gamma_j x_j$ & $A_1$ & $(1/2)^-,(7/2)^-,\ldots$ \\
$(1/2)^+$ & $[P_-]$ & $0^+,1^+$ & $1 \ \ , \ \ \gamma_j x_j$ & $A_1$ & $(1/2)^+,(7/2)^+,\ldots$ \\
 & & & & & \vspace{-0.40cm} \\
\hline
 & & & & & \vspace{-0.40cm} \\
$(3/2)^+$ & $[P_+]$ & $1^+,2^+$ & $\gamma_1 x_1 - \gamma_2 x_2 \ \ $(and cyclic) & $E$ & $(3/2)^+,(5/2)^+,\ldots$ \\
$(3/2)^-$ & $[D_-]$ & $1^-,2^-$ & $\gamma_5 (\gamma_1 x_1 - \gamma_2 x_2) \ \ $(and cyclic) & $E$ & $(3/2)^-,(5/2)^-,\ldots$ \\
 & & & & & \vspace{-0.40cm} \\
\hline
 & & & & & \vspace{-0.40cm} \\
$(5/2)^-$ & $[D_+]$ & $2^-,3^-$ & $\gamma_1 x_2 x_3 + \gamma_2 x_3 x_1 + \gamma_3 x_1 x_2$ & $A_2$ & $(5/2)^-,(7/2)^-,\ldots$ \\
$(5/2)^+$ & $[F_-]$ & $2^+,3^+$ & $\gamma_5 (\gamma_1 x_2 x_3 + \gamma_2 x_3 x_1 + \gamma_3 x_1 x_2)$ & $A_2$ & $(5/2)^+,(7/2)^+,\ldots$\vspace{-0.40cm} \\
 & & & & & \\
\hline
\end{tabular}

\caption{\label{TAB001}static-light meson creation operators.}
\end{center}
\end{table}

When constructing lattice versions of the operators (\ref{EQN001}), one has to replace the integration over the unit sphere by a discrete sum over lattice sites with fixed distance $r$ from the static antiquark. For $J = 1/2$ and $J = 3/2$ we use six lattice sites, i.e.\
\begin{eqnarray}
\label{EQN002} \mathcal{O}^{(J^\mathcal{P})}(\mathbf{x}) \ \ = \ \ \bar{Q}(\mathbf{x}) \sum_{\mathbf{n} = \pm \hat{\mathbf{e}}_1 , \pm \hat{\mathbf{e}}_2 , \pm \hat{\mathbf{e}}_3} \Gamma^{(J^\mathcal{P})}(\hat{\mathbf{n}}) U(\mathbf{x};\mathbf{x}+r \hat{\mathbf{n}}) \psi(\mathbf{x}+r \hat{\mathbf{n}}) ,
\end{eqnarray} 
and for $J = 5/2$ we use eight lattice sites, i.e.\
\begin{eqnarray}
\label{EQN003} \mathcal{O}^{(J^\mathcal{P})}(\mathbf{x}) \ \ = \ \ \bar{Q}(\mathbf{x}) \sum_{\mathbf{n} = \pm \hat{\mathbf{e}}_1 \pm \hat{\mathbf{e}}_2 \pm \hat{\mathbf{e}}_3} \Gamma^{(J^\mathcal{P})}(\hat{\mathbf{n}}) U(\mathbf{x};\mathbf{x}+r \hat{\mathbf{n}}) \psi(\mathbf{x}+r \hat{\mathbf{n}}) .
\end{eqnarray} 
The spatial parallel transporters $U$ in (\ref{EQN002}) are straight paths of links, while in (\ref{EQN003}) we use ``diagonal links'', which are averages over the six possible paths around a cube projected back to SU(3). The states created by these lattice meson creation operators do not form irreducible representations of the rotation group $\mathrm{SO}(3)$, but representations of its cubic subgroup $\mathrm{O}_\mathrm{h}$. Therefore, these states have no well defined total angular momentum $J$. They are linear superpositions of an infinite number of total angular momentum eigenstates. The common notation of the corresponding $\mathrm{O}_\mathrm{h}$ representations together with their angular momentum content are also listed in Table~\ref{TAB001}. Note that we do not consider $\mathrm{O}_\mathrm{h}$ representations $T_1$ and $T_2$, because these representations yield correlation functions, which are numerically identical to those obtained from the operators listed.

% ********************
% ********************
% ********************
% ********************
% ********************
% ********************

\section{Simulation setup}

We use $24^3 \times 48$ gauge configurations produced by the European Twisted Mass Collaboration (ETMC). The fermion action is $N_f = 2$ Wilson twisted mass \cite{Frezzotti:2000nk,Frezzotti:2003ni,Shindler:2007vp} and the gauge action is tree-level Symanzik improved \cite{Weisz:1982zw} with $\beta = 3.9$ corresponding to a lattice spacing $a = 0.0855(5) \, \textrm{fm}$. We consider five different values of the twisted mass $\mu_\mathrm{q}$ (cf.\ Table~\ref{TAB002}). Tuning to maximal twist has been performed at the lightest $\mu_\mathrm{q}$ value yielding $\kappa_\mathrm{cr} = 0.160856$ and ``pion masses'' in the range $300 \, \textrm{MeV} \ltapprox m_\mathrm{PS} \ltapprox 600 \, \textrm{MeV}$. For details regarding this setup we refer to \cite{Boucaud:2007uk,Urbach:2007rt,Boucaud:2008xu}.

% hbar c = 1 = 197.327 MeV fm

% a m_pi(mu = 0.0040) = 0.13587(68) --> 314(2) MeV
% a m_pi(mu = 0.0064) = 0.16937(36) --> 391(1) MeV
% a m_pi(mu = 0.0085) = 0.19403(50) --> 448(1) MeV
% a m_pi(mu = 0.0100) = 0.21004(52) --> 485(1) MeV
% a m_pi(mu = 0.0150) = 0.25864(70) --> 597(2) MeV

\begin{table}[h!]
\begin{center}
\begin{tabular}{|c|c|c|}
\hline
 & & \vspace{-0.40cm} \\
$\mu_\mathrm{q}$ & $m_\mathrm{PS}$ in MeV & number of gauge configurations \\
 & & \vspace{-0.40cm} \\
\hline
 & & \vspace{-0.40cm} \\
$0.0040$ & $314(2)$ & $1400$ \\
$0.0064$ & $391(1)$ & $1450$ \\
$0.0085$ & $448(1)$ & $1350$ \\
$0.0100$ & $485(1)$ & $350$ \\
$0.0150$ & $597(2)$ & $500$\vspace{-0.40cm} \\
 & & \\
\hline
\end{tabular}
\caption{\label{TAB002}$\mu_\mathrm{q}$ values and corresponding pion masses $m_\mathrm{PS}$.}
\end{center}
\end{table}

% ********************
% ********************
% ********************
% ********************
% ********************
% ********************

\section{The static-light meson spectrum}

We determine the static-light meson spectrum from $6 \times 6$ correlation matrices. These matrices are built from trial states, which are generated by operators from Table~\ref{TAB001} belonging to a fixed $\mathrm{O}_\mathrm{h}$ representation. Due to parity breaking discretization errors of the Wilson twisted mass formulation we use states with $\mathcal{P} = +$ and states with $\mathcal{P} = -$ in the same correlation matrix.

To reduce statistical noise and to simplify quark smearing, we use stochastic propagators obtained by inverting four random spin diluted timeslice sources per gauge configuration \cite{Boucaud:2008xu}. We use the HYP2 static action \cite{Hasenfratz:2001hp,Della Morte:2005yc} and apply Gaussian smearing \cite{Gusken:1989qx,Alexandrou:2008tn} to the dynamical quark operators with APE smeared spatial links \cite{Albanese:1987ds}. For details regarding the construction and the computation of these correlation matrices we refer to \cite{sl_paper}.

\begin{figure}[h!]
\begin{center}
\input{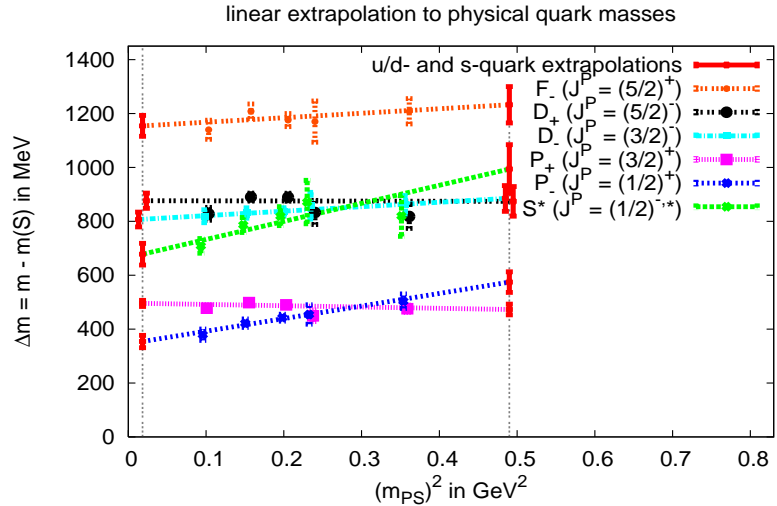}
\caption{\label{FIG001}static-light mass differences linearly extrapolated to the $u/d$ quark mass and the $s$ quark mass.}
\end{center}
\end{figure}

Static-light meson masses are determined both by solving a generalized eigenvalue problem and computing effective masses, and by fitting a suitable ansatz of exponentials to the correlation matrices. Results obtained from both approaches are consistent. Note that static-light meson masses diverge in the continuum limit due to the self energy of the static quark. Therefore, we always consider mass differences of static light mesons, where this self energy exactly cancels, and which are physically meaningful in the continuum limit. Mass differences between various states and the lightest static-light meson ($J^\mathcal{P} = (1/2)^-$ ground state) are shown in Figure~\ref{FIG001} as functions of $(m_\mathrm{PS})^2$.

We extrapolate linearly in $(m_\mathrm{PS})^2$ to the physical $u/d$ quark mass ($m_\mathrm{PS} = 135 \, \textrm{MeV}$) as well as to the physical $s$ quark mass (taken here as $m_\mathrm{PS} = 700 \, \textrm{MeV}$). Note that we consider the unitary sector in both cases, where valence quarks and sea quarks are of the same mass. This implies for the $s$ quark extrapolation a sea of two degenerate $s$ instead of a sea of $u$ and $d$. We plan to improve these computations by using $N_f = 2+1+1$ flavor gauge configurations currently produced by ETMC \cite{Chiarappa:2006ae}. Results are shown in Figure~\ref{FIG001} and Table~\ref{TAB003} together with the corresponding $\chi^2 / \textrm{dof}$ indicating that a straight line is a suitable ansatz.

\begin{table}[h!]
\begin{center}
\begin{tabular}{|lc|c|c|c|}
\hline
 & & & & \vspace{-0.40cm} \\
 & & $u/d$ quark extrapolation: & $s$ quark extrapolation: & \\
\multicolumn{2}{|c|}{$J^\mathcal{P}$} & $m(J^\mathcal{P}) - m(S)$ in MeV & $m(J^\mathcal{P}) - m(S)$ in MeV & $\chi^2 / \textrm{dof}$ \\
 & & & & \vspace{-0.40cm} \\
\hline
 & & & & \vspace{-0.40cm} \\
$(1/2)^{-,\ast}$ & $[S^\ast]$ & $678(40)$ & $994(90)$ & $1.01$ \\
$(1/2)^+$ & $[P_-]$ & $354(23)$ & $574(38)$ & $0.20$ \\
$(3/2)^+$ & $[P_+]$ & $495(11)$ & $473(20)$ & $2.05$ \\
$(3/2)^-$ & $[D_-]$ & $807(27)$ & $885(48)$ & $0.04$ \\
$(5/2)^-$ & $[D_+]$ & $876(28)$ & $874(55)$ & $2.81$ \\
$(5/2)^+$ & $[F_-]$ & $1154(39)$ & $1232(67)$ & $0.72$\vspace{-0.40cm} \\
 & & & & \\
\hline
\end{tabular}
\caption{\label{TAB003}static-light mass differences linearly extrapolated to the $u/d$ quark mass and the $s$ quark mass.}
\end{center}
\end{table}

Performing a similar $u/d$ extrapolation for the $P$ wave mass difference yields \\ $m(P_+) - m(P_-) = 140(22) \, \textrm{MeV}$, i.e.\ the $(1/2)^+$ state is lighter than the $(3/2)^+$ state as is usually expected. We see evidence for a reversal of this level ordering as the light-quark mass increases obtaining a difference $m(P_-) - m(P_+) = 98(33) \, \textrm{MeV}$ at the strange quark mass. It will be interesting to check this  in the continuum limit.

% ********************
% ********************
% ********************
% ********************
% ********************
% ********************

\section{Predictions for $B$ and $B_s$ meson masses}

\begin{figure}[h!]
\begin{center}
\input{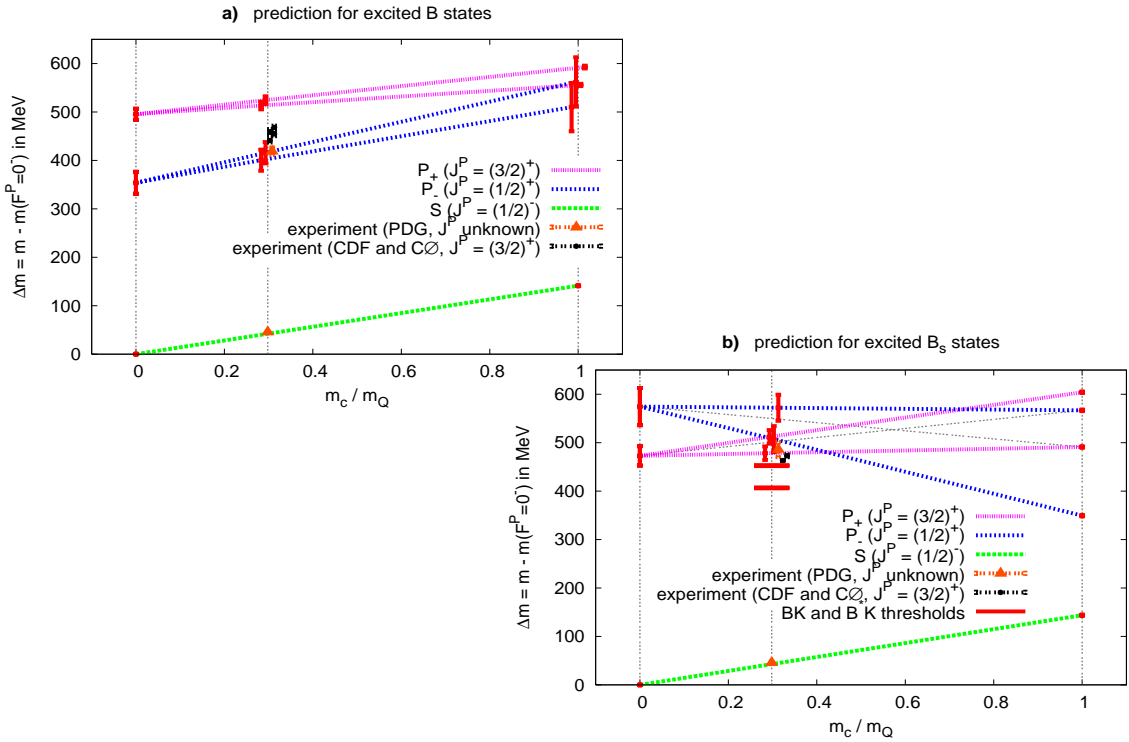}
\caption{\label{FIG002}static-light mass differences linearly interpolated to a heavy $b$ quark.
\textbf{a)}~$B$ mesons.
\textbf{b)}~$B_s$ mesons.
}
\end{center}
\end{figure}

To make predictions regarding the spectrum of $B$ and $B_s$ mesons, we interpolate between our static-light lattice results and experimental data for charmed mesons \cite{PDG}. To this end we assume a linear dependence in $1 / m_Q$, where $m_Q$ is the mass of the heavy quark. Results for $P$ wave $B$ and $B_s$ states are shown in Figure~\ref{FIG002} and Table~\ref{TAB004}. Note that the lines labeled by ``$S (J^\mathcal{P} = (1/2)^-)$'' in Figure~\ref{FIG002} together with the experimental values for $B^\ast$ and $B_s^\ast$ (the triangles intersected by these lines) indicate that straight lines are suited for interpolation.

In Table~\ref{TAB004} we compare our lattice results to experimental data available for $P$ wave $B$ and $B_s$ mesons \cite{PDG,Mommsen:2006ai,Abazov:2007vq,:2007tr,:2007sna}. The lattice results for $B$ states are larger by around $15 \%$ than their experimental counterparts, while there is significantly better agreement for $B_s$ mesons. For a conclusive comparison it will be necessary to investigate the continuum limit, which amounts to considering other values for the lattice spacing.

\begin{table}[h!]
\begin{center}
\begin{tabular}{|c|c|c|c|c||c|c|c|c|c|}
\hline
 & \multicolumn{4}{c||}{\vspace{-0.50cm}} & & \multicolumn{4}{c|}{} \\
 & \multicolumn{4}{c||}{$m-m(F^\mathcal{P}=0^-)$ in MeV} & & \multicolumn{4}{c|}{$m-m(F^\mathcal{P}=0^-)$ in MeV} \\
 & \multicolumn{4}{c||}{\vspace{-0.50cm}} & & \multicolumn{4}{c|}{} \\
\hline
 & & & & & & & & & \vspace{-0.50cm} \\
 state & lattice & PDG$^\ast$ & CDF & C{\O} & state & lattice & PDG$^\ast$ & CDF & C{\O} \\
 & & & & & & & & & \vspace{-0.50cm} \\
\hline
 & & & & & & & & & \vspace{-0.50cm} \\
 $B_0^\ast$ & $400(22)$ & & - & - & $B_{s0}^\ast$ & $507(27)$ & & - & - \\
 $B_1^\ast$ & $416(22)$ & $\uparrow$ & - & - & $B_{s1}^\ast$ & $572(27)$ & $\uparrow$ & - & - \\
 $B_1$ & $513(8)$ &$419(8)$& $455(5)$ & $441(4)$ & $B_{s1}$ & $478(14)$ & $486(16)$ & $463(1)$ & - \\
 $B_2^\ast$ & $524(8)$ & $\downarrow$ & $459(6)$ & $468(4)$ & $B_{s2}^\ast$ & $512(14)$ & $\downarrow$ & $473(3)$ & $473(3)$ \vspace{-0.50cm} \\
 & & & & & & & & & \\
\hline
\end{tabular}
\caption{\label{TAB004}lattice and experimental results for $P$ wave $B$ and $B_s$ states ($^\ast$: $J^\mathcal{P}$ unknown).}
\end{center}
\end{table}

In Figure~\ref{FIG002}b we also plot the $B K$ and $B^\ast K$ threshold ($406 \, \textrm{MeV}$ and $452 \, \textrm{MeV}$ respectively). Our $P$ wave $B_s$ results indicate that the corresponding decays are energetically allowed. Consequently, one can expect that these states have a rather large width compared to e.g.\ certain excited $D_s$ mesons.

% ********************
% ********************
% ********************
% ********************
% ********************
% ********************

\section{Conclusions}

We have studied the static-light meson spectrum by means of $N_f = 2$ Wilson twisted mass lattice QCD. We have considered five different values of the dynamical quark mass corresponding to $300 \, \textrm{MeV} \ltapprox m_\mathrm{PS} \ltapprox 600 \, \textrm{MeV}$. We have performed an extrapolation in the light quark mass to the physical $u/d$ mass and $s$ mass respectively, as well as an interpolation in the heavy quark mass to the physical $b$ mass. Our results agree within $\ltapprox 15 \%$ with currently available experimental results for $P$ wave $B$ and $B_s$ states.

Future plans regarding this project include an investigation of the continuum limit, which amounts to considering other values for the lattice spacing. Moreover, we plan to perform similar computations on $N_f = 2+1+1$ flavor gauge configurations currently produced by ETMC. In particular, this will allow us to include a sea of $u$ and $d$ in $B_s$ computations. We also intend to determine the static-light decay constants $f_B$ and $f_{B_s}$.

% ********************
% ********************
% ********************
% ********************
% ********************
% ********************

\begin{acknowledgments}

MW would like to thank Carsten Urbach for help in retrieving and handling ETMC gauge configurations. Moreover, we acknowledge useful discussions with Benoit Blossier and Carsten Urbach. This work has been supported in part by the DFG Sonderforschungsbereich/Transregio SFB/TR9-03.

\end{acknowledgments}

% ********************
% ********************
% ********************
% ********************
% ********************
% ********************

% ********************
% ********************
% ********************
% ********************
% ********************

\end{document}